\documentclass[preprint,12pt]{elsarticle}

\usepackage{amssymb}

\usepackage{latexsym}
\usepackage{graphicx,amssymb,psfrag}
\usepackage{amsmath}
\usepackage{enumerate}
\usepackage{varioref}
\usepackage{rotating}
\usepackage{graphics}
\usepackage[dvips]{epsfig}
\usepackage{amssymb}
\usepackage{url}

\def\boxit#1{\vbox{\hrule\hbox{\vrule\kern4pt
  \vbox{\kern1pt#1\kern1pt}
\kern2pt\vrule}\hrule}}

\newcommand\nc{\newcommand}

\newtheorem{observation}{\bfseries Observation}
\newtheorem{theorem}{\bfseries Theorem}
\newtheorem{lemma}{\bfseries Lemma}
\newtheorem{rull}{\bfseries Rule}
\newtheorem{corollary}{\bfseries Corollary}
\newtheorem{definition}{\bfseries Definition}

\nc{\crl}[2]{\begin{corollary}\label{crl:#1} #2 \end{corollary}}
\nc{\dfn}[2]{\begin{definition}\label{def:#1} #2 \end{definition}}
\nc{\llem}[2]{\begin{lemma}\label{lem:#1} #2 \end{lemma}}
\nc{\thmm}[2]{\begin{theorem}\label{thm:#1} #2\end{theorem}}
\nc{\rul}[2]{\begin{rull}\label{rull:#1} #2\end{rull}}

\nc{\eqn}[2]{\begin{eqnarray}\label{eqn:#1} #2 \end{eqnarray}}

\nc{\fig}[4]{\begin{figure}[h]
\begin{center}
\includegraphics[width=#2\textwidth]{#4}
\end{center}
\caption{#3}\label{fig:#1}
\end{figure}}

\nc{\tbl}[3]{\begin{table}[hbt] #3 \caption{#2} \label{tab:#1}
\end{table}}

\nc{\refc}[1]{Corollary~\ref{crl:#1}}
\nc{\refd}[1]{Definition~\ref{def:#1}}
\nc{\reff}[1]{Figure~\ref{fig:#1}}
\nc{\refl}[1]{Lemma~\ref{lem:#1}}
\nc{\refp}[1]{Proposition~\ref{prp:#1}}
\nc{\reft}[1]{Theorem~\ref{thm:#1}} \nc{\refe}[1]{(\ref{eqn:#1})}
\nc{\reftb}[1]{Table~\ref{tab:#1}}
\nc{\refr}[1]{Rule~\ref{rull:#1}}

\nc{\reffc}[1]{Fact~\ref{fact:#1}}

\nc{\pff}[1]{ \noindent \emph{Proof.} #1 \hfill \qed\par}

\long\def\invis#1{}

\begin{document}

\begin{frontmatter}



\title{On a generalization of Nemhauser and Trotter's local optimization theorem}


\author{Mingyu Xiao}
\ead{myxiao@uestc.edu.cn}
\address{School of Computer Science and Engineering,\\
University of Electronic Science and Technology of China, China
}

\begin{abstract}

Fellows, Guo, Moser and Niedermeier~[JCSS2011] proved a generalization of Nemhauser and Trotter's theorem, which applies to \textsc{Bounded-Degree Vertex Deletion}
(for a fixed integer $d\geq 0$,  to delete $k$ vertices of the input graph to make
the maximum degree of it $\leq d$)
and gets a linear-vertex kernel for
$d=0$ and $1$, and a superlinear-vertex kernel for each $d\geq 2$.
It is still left as an open problem whether \textsc{Bounded-Degree Vertex Deletion} admits a linear-vertex kernel for each $d\geq 3$.
In this paper, we refine the generalized Nemhauser and Trotter's theorem and get a linear-vertex kernel for each $d\geq 0$.

\end{abstract}


\end{frontmatter}

\section{Introduction}
\textsc{Vertex Cover}, to find a minimum set of vertices in a graph such that each edge in the graph is incident on at least one vertex in this set, is one of the most fundamental problems in graph algorithms, graph theory, parameterized algorithms, theories of NP-completeness
and many others.
Nemhauser and Trotter~\cite{NT-theorem}
proved a famous theorem (NT-Theorem) for  \textsc{Vertex Cover}.
\thmm{ntthm}{ \emph{[\textbf{NT-Theorem}]}
For an undirected graph $G=(V,E)$ of $n=|V|$ vertices and $m=|E|$ edges, there is an $O(\sqrt{n}m)$-time
algorithm to compute two disjoint vertex subsets $C$
and $I$ of $G$ such that for any minimum vertex cover $K'$ of the induced subgraph $G[V \setminus (C \cup I)]$, $K' \cup C$ is
a minimum vertex cover of $G$ and
$$|K'|\geq {\frac{|V \setminus (C \cup I)|}{2}}.$$
}
This theorem provides a polynomial-time algorithm to reduce the size of the input
graph by possibly finding partial solution.
It turns out that NT-Theorem has great applications in approximation algorithms~\cite{BE:WVC,hochbaum,khuller} and parameterized algorithms~\cite{CKJ:VC,A:crown2}. We can see that $V\setminus I$ is a 2-approximation solution
and $G[V \setminus (C \cup I)]$ is a $2k$-vertex kernel
of the problem taking the size of
the solution as the parameter $k$. Lokshtanov et al.~\cite{lp} also apply NT-Theorem to branching algorithms for \textsc{Vertex Cover} and some other related problems.
Due to NT-Theorem's practical usefulness and theoretical depth in graph theory, it has attracted numerous
further studies and follow-up work~\cite{FG:gNT,BRH:extension,CC:WVC,A:crown2}.
Bar-Yehuda, Rawitz and Hermelin~\cite{BRH:extension} extended NT-Theorem for a generalized vertex cover problem, where edges are allowed not to be covered at a certain
predetermined penalty. Fellows, Guo, Moser and Niedermeier~\cite{FG:gNT} extended NT-Theorem for \textsc{Bounded-Degree Vertex Deletion}.

In this paper, we are interested in \textsc{Bounded-Degree Vertex Deletion}.
A \emph{$d$-degree deletion set} of a graph $G$ is a subset of vertices, whose deletion
leaves a graph of maximum degree at most $d$. For each fixed $d$, \textsc{Bounded-Degree Vertex Deletion} is to find
a $d$-degree deletion set of minimum size in an input graph.
\textsc{Bounded-Degree Vertex Deletion} and its ``dual problem'' to find maximum $s$-plexes have applications in computational biology~\cite{FG:gNT,CF:copath}
and social network analysis~\cite{SF:plex,BBH:plex}.
There is a substantial amount of theoretical work on this problem~\cite{khmn,NRT,SF:plex}, specially in parameterized complexity~\cite{BBNU:treewidth,FG:gNT,CF:copath}.

Since \textsc{Vertex Cover} is a special case of \textsc{Bounded-Degree Vertex Deletion},
we are interested in finding a local optimization theorem similar to NT-Theorem for \textsc{Bounded-Degree Vertex Deletion}.
Fellows, Guo, Moser and Niedermeier~\cite{FG:gNT} made a great progress toward to this interesting problem by giving the following theorem.

\thmm{ntg-thm}{ \emph{\cite{FG:gNT}}
For an undirected graph $G=(V,E)$ of $n=|V|$ vertices and $m=|E|$ edges, any constant $\varepsilon > 0$ and any integer $d\geq0$, there is an $O(n^4m)$-time
algorithm to compute two disjoint vertex subsets $C$
and $I$ of $G$ such that for any minimum $d$-degree deletion set $K'$ of the induced subgraph $G[V \setminus (C \cup I)]$, $K' \cup C$ is
a minimum $d$-degree deletion set of $G$, and

\[|K'|\geq {\frac{|V \setminus (C \cup I)|}{d^3+4d^2+6d+4}}~~~~~~\mbox{for}~~d\leq 1, ~~~~~~~\mbox{and}\]
\[|K'|^{1+\varepsilon}\geq {\frac{|V \setminus (C \cup I)|}{c}}~~~~~~~\mbox{for}~~d\geq 2, ~~~~~~~~~~~~\]
where $c$ is a function of $d$ and $\varepsilon$.
}

In this theorem, for $d\geq 2$, the number of remaining vertices in $V \setminus (C \cup I)$
is not bounded by a constant times of the solution size $|K'|$ of $G[V \setminus (C \cup I)]$.
This is a significant difference between this theorem and the NT-Theorem for \textsc{Vertex Cover}.
In terms of parameterized algorithms, \reft{ntg-thm} cannot get a linear-vertex kernel for
\textsc{Parameterized Bounded-Degree Vertex Deletion} (with parameter $k$ being the solution size) for  each $d\geq 2$.
In fact, in an initial version \cite{FG:gNTc} of Fellows, Guo, Moser and Niedermeier's paper, a better result was claimed, which can get
a linear-vertex kernel for \textsc{Parameterized Bounded-Degree Vertex Deletion} for  each $d\geq 0$.
Unfortunately, the proof in \cite{FG:gNTc} is incomplete.
We also note that Chen et al.~\cite{CF:copath} proved a $37k$-vertex kernel for \textsc{Bounded-Degree Vertex Deletion} for $d=2$.
However, whether \textsc{Bounded-Degree Vertex Deletion} for each $d\geq 3$ allows a linear-vertex kernel is not known.
In this paper, based on Fellows, Guo, Moser and Niedermeier's work \cite{FG:gNTc},
we close the above gap by proving the following theorem for \textsc{Bounded-Degree Vertex Deletion}.

\thmm{our-thm}{\emph{\textbf{[Our result]}}
For an undirected graph $G=(V,E)$ of $n=|V|$ vertices and $m=|E|$ edges and any integer $d\geq 0$, there is an $O(n^{5/2}m)$-time
algorithm to compute two disjoint vertex subsets $C$
and $I$ of $G$ such that for any minimum $d$-degree deletion set $K'$ of the induced subgraph $G[V \setminus (C \cup I)]$, $K' \cup C$ is
a minimum $d$-degree deletion set of $G$ and

\[|K'|\geq {\frac{|V \setminus (C \cup I)|}{d^3+4d^2+5d+3}}.\]
}

From this version of the generalized Nemhauser and Trotter's theorem, we can get a
$(d^3+4d^2+5d+3)k$-vertex kernel for
\textsc{Bounded-Degree Vertex Deletion} parameterized by the size $k$ of the solution,
which is linear in $k$ for any constant $d\geq 0$. There is no difference between the cases that $d\leq 1$ and $d\geq 2$ anymore.
%
%
For the special case that $d=0$, our theorem specializes a $3k$-vertex kernel for \textsc{Vertex Cover}, while
\reft{ntg-thm} provides a $4k$-vertex kernel and NT-Theorem provides a $2k$-vertex kernel.
For the special case that $d=1$, our theorem provides a $13k$-vertex kernel and \reft{ntg-thm} provides a $15k$-vertex kernel.
For the special case that $d=2$, our theorem obtains a $37k$-vertex kernel, the same result obtained by Chen et al.~\cite{CF:copath}.

Recently, Dell and van Melkebeek~\cite{dell} showed that unless the polynomial-time hierarchy collapses,
\textsc{Parameterized Bounded-Degree Vertex Deletion}
does not have kernels consisting of $O(k^{2-\epsilon})$ edges for any constant $\epsilon>0$, which implies that linear size would be the best possible bound on the number of vertices in any kernel for this problem. It has also been proved by
Fellows, Guo, Moser and Niedermeier~\cite{FG:gNT} that when $d$ is not bounded, \textsc{Parameterized Bounded-Degree Vertex Deletion} is W[2]-hard. Then unless FPT=W[2], it is impossible to remove $d$ from the size function of any kernel of this problem. These two hardness results also imply  that our result is `tight' in some sense.

The framework of our algorithm follows that of Fellows, Guo, Moser and Niedermeier's algorithm~\cite{FG:gNT}.
But we still need some new and nontrivial ideas to get our result.
For the purpose of presentation, we will define a decomposition, called `$d$-bounded decomposition' to
prove \reft{our-thm} and construct our algorithms. This decomposition can be regarded as an extension of the crown decomposition for \textsc{Vertex Cover}~\cite{A:crown,cfj:crown}, but more sophisticated.
To compute $C$ and $I$ in \reft{our-thm}, we will change to compute a proper $d$-bounded decomposition.
Some similar ideas in construction of crown decompositions as in Fellows, Guo, Moser and Niedermeier's algorithm for \reft{ntg-thm}~\cite{FG:gNT} are used to construct our decomposition. The detailed differences between our and previous algorithms will be addressed in Section~\ref{sec_alg}.
Before introducing the decompositions,
 we first give the notation system in this paper.

\section{Notation system}
Let $G=(V,E)$ stand for a simple undirected graph with a set $V$ of $n=|V|$ vertices and a set $E$ of $m=|E|$ edges.
For simplicity,  we may denote a singleton set $\{v\}$   by $v$.
For a vertex subset $V'$, a vertex in $V'$ is denoted by \emph{$V'$-vertex}. The graph induced by $V'$ is denoted by $G[V']$. We also use $N(V')$ to denote the set of vertices in $V\setminus V'$ adjacent to some vertices in $V'$ and let $N[V']=N(V')\cup V'$.
The vertex set and edge set of a graph $G'$ are denoted by $V(G')$ and $E(G')$, respectively.
A bipartite graph with two parts of vertices $A$ and $B$ and edge set $E_H$ is denoted by $H=(A, B, E_H)$.

For  an integer $d'\geq 1$, a star with $d'+1$ vertices is called a  {\em $d'$-star}.
For $d'>1$, the unique vertex of degree $>1$ in a $d'$-star is called the \emph{center}
of the star and all other degree-1 vertices are called the \emph{leaves} of the star.
For a 1-star, any vertex can be regarded as a \emph{center} and the other vertex as a \emph{leaf}.
A star with a center $v$ is also called a star \emph{centered at} $v$.
For two disjoint vertex sets $V_1$ and $V_2$, a set of stars is \emph{from $V_1$ to $V_2$} if
the centers of the stars are in $V_1$ and leaves are in $V_2$.
A \emph{$_\leq d'$-star} is a star with at most $d'$ leaves.
A \emph{$d'$-star packing} (resp., \emph{$_\leq d'$-star packing}) is
a set of vertex-disjoint $d'$-stars (resp., $_\leq d'$-stars).

For each $d\geq 0$, a \emph{$d$-degree deletion set} of a graph is a subset of vertices the deletion of which makes
the maximum degree of the remaining graph at most $d$.
We use $\alpha(G)$ to denote the size of a minimum $d$-degree deletion set of a graph $G$.

Next, we introduce the decomposition techniques in Section~\ref{sec_dec} and then describe and analyze our
algorithms in Section~\ref{sec_alg}.

\section{The decomposition
techniques}\label{sec_dec}
Crown decomposition is a powerful tool to obtain kernels for \textsc{Vertex Cover}.
This technique was firstly introduced in~\cite{A:crown} and~\cite{cfj:crown} and found to be very useful
in designing kernelization algorithms for \textsc{Vertex Cover} and related problems~\cite{A:crown2,CC:WVC,X:vc3}.

\begin{figure}[h]
\begin{center}
\includegraphics[width=0.9\textwidth]{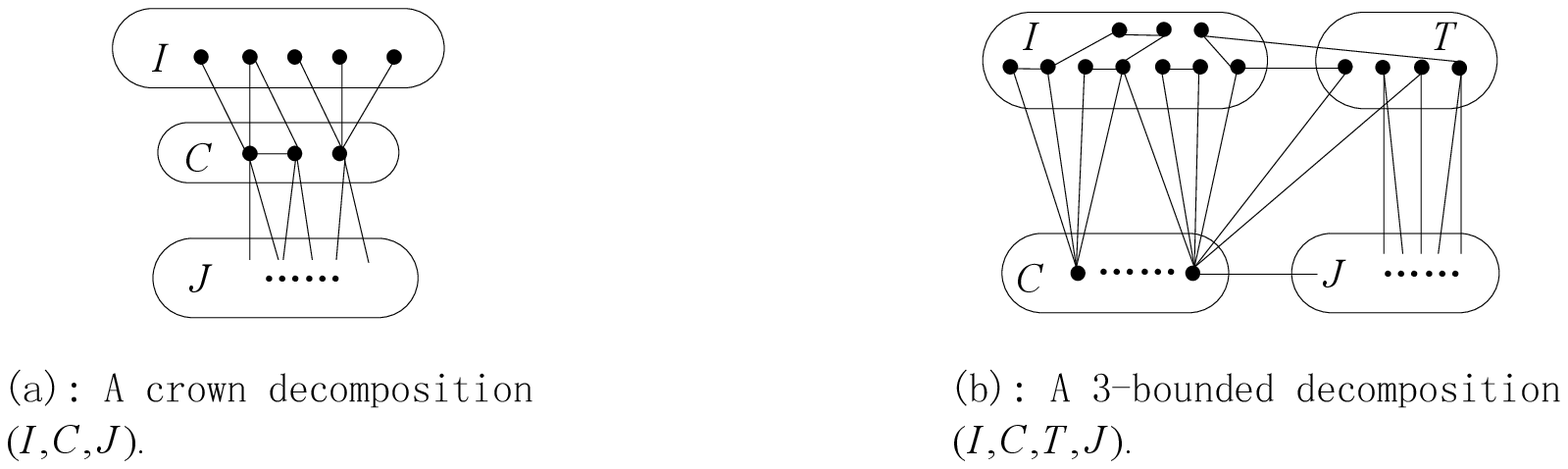}
\end{center}
\caption{Decompositions}\label{fig1}
\end{figure}

\dfn{crown}{\emph{[\textbf{Crown Decomposition}]} A crown decomposition of a graph $G$ is
a partition of the vertex set of $G$ into three sets $I$, $C$ and
$J$ such that\\
(1) $I$  is an independent set, \\
(2) there are no edges between $I$ and $J$, and \\
(3) there is a matching $M$ on the edges between $I$ and $C$ such that all vertices in $C$ are matched.
}
See Figure~\ref{fig1}(a) for an illustration for crown decompositions. In some references, $I\neq \emptyset$ is also required in the definition of crown decompositions. Here we allow $I=\emptyset$ for the
purpose of presentation.
It is known that
\llem{crown_local}{\emph{\cite{A:crown}} Let $(I,C,J)$ be a crown decomposition of $G$. Then $(I,C)$ satisfies the local optimality condition in \reft{ntthm}, i.e.,
$K' \cup C$ is a minimum vertex cover of $G$ for any minimum vertex cover $K'$ of the induced subgraph $G[V \setminus (I \cup C)]$.
}
By this lemma, we can reduce the instance of \textsc{Vertex Cover} by removing $I\cup C$ of a  crown decomposition. There are some methods that find certain crown decompositions of a graph and result in a linear-vertex
kernel for \textsc{Vertex Cover}~\cite{A:crown2}.

In this paper, we will use \emph{$d$-bounded decomposition}, which extends the definition of crown decompositions and \refl{crown_local}.
Let $A$ and $B$ be two disjoint vertex subsets of a graph $G$. A \emph{full $d'$-star packing from $A$ to $B$} is a set of $|A|$ vertex-disjoint $d'$-stars
with centers in $A$ and leaves in $B$. The third item in \refd{crown} means that there is a full $1$-star packing from $C$ to $I$. We define the following decomposition.

\dfn{general_dc}{\emph{[\textbf{$d$-Bounded Decomposition}]} A $d$-bounded decomposition of a graph $G=(V,E)$ is
a partition of the vertex set of $G$ into four sets $I$, $C$, $T$ and
$J$ such that\\
(1) any vertex in $I\cup T$ is of degree $\leq d$ in the induced subgraph $G[V\setminus C]$, \\
(2) there are no edges between $I$ and $J$, and \\
(3) there is a full $(d+1)$-star packing from $C$ to $I$.
}

An illustration for $d$-bounded decompositions is given in Figure~\ref{fig1}(b).
We have the following \refl{dc_local} for $d$-bounded decompositions.
This lemma can be derived from the lemmas in \cite{FG:gNT}, although $d$-bounded decomposition is not formally defined
in \cite{FG:gNT}.

\llem{dc_local}{Let $(I,C,T,J)$ be a $d$-bounded decomposition of $G$. Then $(I,C)$ satisfies the local optimality condition in \reft{our-thm}, i.e.,
$K' \cup C$ is a minimum $d$-degree deletion set of $G$ for any minimum $d$-degree deletion set $K'$ of the induced subgraph $G[V \setminus (I \cup C)]$.
}
\pff{ First, we show that $K' \cup C$ is a $d$-degree deletion set of $G$.
If there is vertex $v_0$ of degree $\geq d+1$ in $G[V \setminus (K' \cup C)]$, then $v_0$ should be a $J$-vertex, since
any vertex in $I\cup T$ is of degree $\leq d$ after removing $C$ by the definition of the decomposition.
Note that no $J$-vertex is adjacent to an $I$-vertex.
Then $v_0$ would also be a vertex of degree $\geq d+1$ in $G[V \setminus (K' \cup C\cup I)]$,
which implies a contradiction that $K'$ is not a $d$-degree deletion set of $G[V \setminus (I \cup C)]$.
So no vertex of degree $\geq d+1$ exists in $G[V \setminus (K' \cup C)]$.

Next, we prove the minimality.
Let $D$ be an arbitrary minimum $d$-degree deletion set of $G$.
Let $D_1=D\cap (I\cup C)$ and $D_2=D\cap (T\cup J)$.
Since there is a full $(d+1)$-star packing from $C$ to $I$, we know that any $d$-degree deletion set contains
at least $|C|$ vertices in the $(d+1)$-star packing. So we have that
$$|D_1|\geq |C|.$$
Set $D_2$ is a $d$-degree deletion set of $G[V\setminus D_1]$ and
set $K'$ is a minimum $d$-degree deletion set of $G[V \setminus (I \cup C)]$.
Note that $D_1 \subseteq I\cup C$ and then $G[V \setminus (I \cup C)]$ is an induced subgraph of $G[V\setminus D_1]$.
So it holds that
$$|D_2|\geq |K'|.$$
Therefore,  $|K' \cup C|=|K'|+|C|\leq |D_1|+|D_2|=|D|$.
}

By \refl{dc_local}, we can reduce an instance by removing $I\cup C$ if the graph has a $d$-bounded decomposition $(I,C,T,J)$.
This is the main idea how we get \reft{our-thm} and kernels for our problem.
Here arises a problem how to find a $d$-bounded decomposition $(I,C,T,J)$ of a graph such that $I\neq \emptyset$ if it exists.
First, we give a simple observation.
\begin{observation}{\label{simple_dc}
Let $R$ be a set of vertices $v$ such that any vertex in $N[v]$ is of degree $\leq d$.
Then $(I=R, C=\emptyset, T=N(R), J=V\setminus(I\cup T))$ is a $d$-bounded decomposition of $G$.}
\end{observation}

By \refl{dc_local} and Observation~\ref{simple_dc}, we can reduce an instance by removing from the graph the set $B$ of vertices $v$ such that any vertex in $N[v]$ is of degree $\leq d$.
For more general cases, in this paper we will show that
\thmm{normal_dc}{ For a given graph $G=(V,E)$ and an integer $d\geq 0$,
there is a special $d$-degree deletion set $X$ of $G$ with $|X| \leq (d+2)\alpha(G)$ such that if $|V\setminus X| >{\frac{(d+1)(d^2+3d+1)}{d+2}}|X|$,
then $G$ admits a $d$-bounded decomposition $(I,C,T,J)$ with $I\neq \emptyset$. The special $d$-degree deletion set $X$ and
$d$-bounded decomposition $(I,C,T,J)$ can be found in $O(n^{3/2}m)$ time.}

In the next section, we construct an algorithm to prove this theorem.


\section{Algorithms}\label{sec_alg}

We first introduce an algorithm to find $d$-bounded decompositions of graphs, based on which we can easily get an algorithm for
the generalization of NT-theorem in \reft{our-thm}.

\subsection{The algorithm for decompositions}
First of all, we give the main idea of our algorithm to find a $d$-bounded decomposition $(I,C,T,J)$ of a graph $G=(V,E)$. It contains three major phases.
\\
Phase 1: find a partition $(X,Y)$ of the vertex set $V$ such that the maximum degree in $G[Y]$ is at most $d$.\\
Phase 2: find two subsets $C'\subseteq X$ and $I'\subseteq Y$ satisfying \emph{Basic Condition}: there is a full $(d+1)$-star packing from $C'$ to $I'$ and
there is no edge between $I'$ and $X\setminus C'$.\\
Phase 3: iteratively move some vertices out of $I'$ and some vertices out of $C'$ to make $(I',C',T'=N(I')\setminus C',J'=V\setminus(I'\cup C'\cup T'))$
 a $d$-bounded decomposition.

In fact, the first two phases of our algorithm are almost the same as that of Fellows, Guo, Moser and Niedermeier's algorithm~\cite{FG:gNT}.
However, in Phase 3, our algorithm uses a different method to compute $I'$ and $C'$. This is critical for us to get an improvement.

\medskip
\noindent
\textbf{Phase 1.}
For Phase 1, we can find a maximal $(d+1)$-star packing $S$ and let $X=V(S)$. By the maximality of $S$, we know that $X$ is a $d$-degree deletion set and $G[Y]$ has no vertex of degree $>d$.
Then the partition $(X,Y)$ satisfies the condition in Phase 1. In order to obtain a good performance, our algorithm may not use an arbitrary
maximal $(d+1)$-star packing $S$.
When we obtain a new $(d+1)$-star packing $S'$ such that $|S'|> |S|$ in our algorithm, we will update  $X$ by letting $X=V(S')$.

\medskip
\noindent
\textbf{Phase 2.} After obtaining $(X,Y)$ in Phase 1, our algorithm finds two special sets $C'\subseteq X$ and $I'\subseteq Y$ in Phase 2.
To find $C'$ and $I'$ satisfying Basic Condition, we need to find a special $_{\leq} (d+1)$-star packing from $X$ to $Y$,
which can be computed by the algorithms for finding maximum matchings in bipartite graphs. Note that the idea of computing $_{\leq} (d+1)$-stars from $X$ and $Y$ has been used to solve some other problems in references~\cite{fvs,fomin,cygan}.

We consider the bipartite graph $H=(X,Y,E_H)$ with edge set $E_H$ being the set of edges between $X$ and $Y$ in $G$, and are going to find a $_{\leq} (d+1)$-star  packing from $X$ to $Y$ in $H$.
Note that a $Y$-vertex no adjacent to any vertex in $X$ will become a degree-0 vertex in $H$.
We construct an auxiliary bipartite graph $H'=(X_1\cup X_2\cup \dots X_{d+1},Y,E'_H)$,
where each $X_i$ $(i=1,2,\dots, d+1)$ is a copy of $X$ and a vertex  $v_i\in X_i$ is adjacent
to a vertex $u\in Y$ if and only if the corresponding vertex $v\in X$ is adjacent to $u$ in $H$. For a vertex $v\in X$, we may use $v_i$ to denote its corresponding vertex in $X_i$.

We find a maximum matching $M'$ in $H'$ by using an $O(n^{1/2}m)$-time algorithm~\cite{ET:bm,hk:bm}.
Let $M$ be the set of edges in $H$ corresponding to the matching $M'$, i.e., an edge $uv$ ($u\in Y$ and $v\in X$) of $H$ is in $M$ if and only if $uv_i$ is in $M'$ for some $v_i$ corresponding to $v$. Edges in $M$ are called \emph{marked} and others are called \emph{unmarked}.
Since $M'$ is a matching in $H'$, we have that $|M|=|M'|$.
The set of marked edges in $H$ forms a $_{\leq} (d+1)$-star packing $S_{\leq d+1}$. This is the $_{\leq} (d+1)$-star packing we are seeking for.
It is also easy to observe that

\llem{mm}{Graph $H$ has a $_{\leq} (d+1)$-star packing  containing $t$ edges if and only if $H'$ has a matching of size $t$.}

Next, we analyze some properties of $S_{\leq d+1}$ and find $C'$ and $I'$ satisfying Basic Condition based on these properties.

\begin{figure}[h]
\begin{center}
\includegraphics[width=0.4\textwidth]{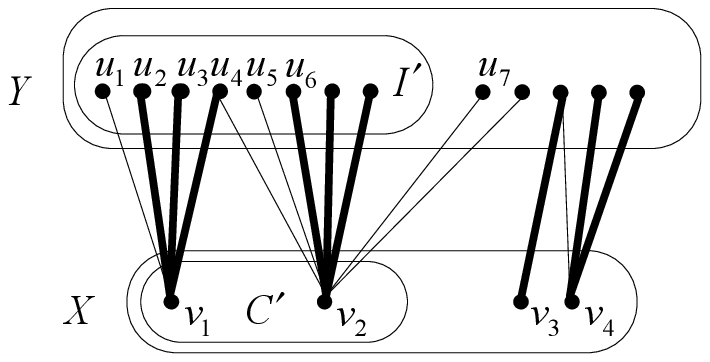}
\end{center}
\caption{An illustration for $I'$ and $C'$, where thick edges are marked edges, $v_1$ and $v_2$
are fully tagged vertices, $u_1$ and $u_5$ are untagged vertices, and $u_1v_1u_4v_2u_6$ is an $M$-alternating
path}\label{fig2}
\end{figure}

Let $S_{d+1}$ denote the set of $(d+1)$-stars in $S_{\leq d+1}$.
An $X$-vertex in a star in $S_{d+1}$ is \emph{fully tagged}. Then $X\cap V(S_{d+1})$ is the set of fully tagged vertices.
A $Y$-vertex is \emph{untagged} if it is adjacent to at least one vertex in $X$ in $H$ but not contained in any star in  $S_{\leq d+1}$.
A path $P$ in $H$ 
that alternates between edges not in $M$ and edges in $M$ is called an \emph{$M$-alternating path}. Please
see Figure~\ref{fig2} for an illustration of these definitions.

\llem{111}{If there is an $M$-alternating path $P$ from an untagged vertex $u\in Y$ to a vertex $v\in X$ in $H$,
then $v$ is fully tagged.
}
\pff{ Note that the edge incident on $u$ in $P$, which can be regarded as the first edge in $P$, is unmarked, and $P$ contains odd number of edges since
$u\in Y$ and $v\in X$.
According to the definition of $M$-alternating paths, we know that $P$ contains more unmarked edges than marked edges.
Replacing $M\cap E(P)$ by $E(P)\setminus M$ in $M$ produces $M_0$.
If $v$ is not fully tagged, then $M_0$ still can form a $_{\leq} (d+1)$-star packing in $H$. By \refl{mm}, there will be a matching of size $|M_0|>|M'|$ in $H'$, contradicting to the maximality of $M'$. So $v$ is fully tagged.
}
\medskip
Next, we are going to set $C'$ and $I'$.
If there is no untagged vertex, let $C'=\emptyset$. Otherwise let $C'$ be the set of $X$-vertices connected with at least one untagged vertex by an $M$-alternating path in $H$. Let $X'=X\setminus C'$.
Let $Y'$ be the set of $Y$-vertices that is a leaf of a  $_{\leq} (d+1)$-star in  $S_{\leq d+1}$ that is centered at a vertex in $X'$, and $I'=Y\setminus Y'$.

\llem{112}{The two sets $C'$ and $I'$ obtained above satisfy Basic Condition.}
\pff{
By the definition of $C'$ and \refl{111}, we know that all vertices in $C'$ are fully tagged. Any leaf of a star centered at a vertex in $C'$ will
not be in $Y'$ since each vertex in $Y$ is in at most one star in $S_{\leq d+1}$.
Then we know that the set of stars in $S_{\leq d+1}$ centered at vertices in $C'$ is a full $(d+1)$-star packing from $C'$ to $I'$.

Next, we show that there is no edge between $I'$ and $X'=X\setminus C'$. Assume to the contrary that there is an edge $uv$ between $I'$ and $X'$,
where $u \in I'$ and $v \in X'$. The vertex $u$ cannot be an untagged vertex, otherwise if $v$ is fully tagged then $v$ would be included to $C'$ by
the definition of $C'$, and
if $v$ is not fully tagged then $uv$ could be added to $M$ to obtain a matching of larger size.
So $u$ is a leaf of a $(d+1)$-star in $S_{\leq d+1}$ centered at a vertex $v_0\in C'$ and $v_0u$ is an $M$-edge in $H$. We can find an $M$-alternating path $P$ from an untagged vertex $u_0$ to $u$ in $H$. There is an $M$-alternating path $P'$ from an untagged vertex $u_0$ to $v_0$ according to the definition of $C'$. If $P'$ passes $u$ then let $P$ be the subpath of $P'$ from $u_0$ to $u$. Otherwise we let $P$ be the path adding $v_0u$ to the end of $P'$.
Then $P$ is an $M$-alternating path from an untagged vertex $u_0$ to $u$.
Let $P^*$ be the path adding $uv$ to the end of $P$. We can see that $P^*$ is still an $M$-alternating path,
which is from an untagged vertex $u_0$ to a $J'$-vertex $v$. However, according to the definition of $C'$, $v$ should be included to $C'$.
For any case, there is a contradiction.

So $C'$ and $I'$ satisfy Basic Condition.
}

\medskip

We describe the above progress to compute $C'$ and $I'$ as an algorithm ${\tt basic}(G,X,Y)$ in Figure~\ref{phase2}, which will be used as a subalgorithm in our main algorithm.
Step~1 of ${\tt basic}(G,X,Y)$ takes linear time and $H'$ has $O(n)$ vertices and $O(m)$ edges since $d$ is a constant.
Step~2 takes $O(n^{1/2}m)$ time to compute a maximum matching $M'$ in the bipartite $H'$.
In Step~3, $C'$ can be computed in linear time by contracting all untagged vertices into a single vertex and using BFS. Therefore,
\llem{time1}{Algorithm ${\tt basic}(G,X,Y)$ runs in $O(n^{1/2}m)$ time.}

\begin{figure*}

\rule{\linewidth}{0.4mm}

\textbf{Input}: A graph $G=(V,E)$ and a partition $(X, Y)$ of the vertex set $V$. \\
\textbf{Output}: Two sets $C'\subseteq X$ and $I'\subseteq Y$ satisfying the Basic Condition.
\begin{enumerate}
\item Compute the bipartite graph $H$ and the auxiliary bipartite graph $H'$.
\item Compute a maximum matching $M'$ in $H'$ and the corresponding edge set $M$ and the $_{\leq} (d+1)$-star packing $S_{\leq d+1}$ in $H$.
\item Let $C'$ be $\emptyset$ if there is no untagged vertex, and the set of $X$-vertices connected with at least one untagged vertex by an $M$-alternating path in $H$ otherwise. Let $X'\leftarrow X\setminus C'$. Let $Y'$ be the set of $Y$-vertices each of which is a leaf of a  $_{\leq} (d+1)$-star centered at a vertex in $X'$ and let $I'\leftarrow Y\setminus Y'$.
\item Return $(C',I')$.

\end{enumerate}

\rule{\linewidth}{0.4mm}
\caption{Algorithm ${\tt basic}(G,X,Y)$}\label{phase2}
\end{figure*}

Note that all untagged vertices will be in $I'$. So if the size of $Y$ is large, for example $|Y|> (d+1)|X|$,
we can guarantee that there is always some untagged vertices and the set $I'$ returned by ${\tt basic}(G,X,Y)$
is not an empty set.

\medskip
\noindent
\textbf{Phase 3.} After obtaining $(C',I')$ from Phase~2, we look at the partition $\mathcal{P}=(I', C', T'=N(I')\setminus C',J'=V\setminus(I'\cup C'\cup T'))$.
Since  there is no edge between $I'$ and $X'=X\setminus C'$, we know that $T'\subseteq Y$ and $X'\subseteq J'$.
Then there is no edge between $I'$ and $J'$. The partition $\mathcal{P}$ satisfies Conditions (2) and (3) in \refd{general_dc} for $d$-bounded decompositions.
Next, we consider Condition (1).
Let $G^*=G[V\setminus C']$. Any vertex in $I'$ is of degree $\leq d$ in $G^*$, because $G[Y]=G[V\setminus X]$ has maximum degree $\leq d$ and $I'$-vertices are not adjacent to any vertex in $X\setminus C'$. Although  $T'=N(I')\setminus C' \subseteq Y$, vertices in $T'$ is possible to be of degree $>d$ in $G^*$.
In fact, we only know that each vertex in $T'$ is of degree $\leq d$ in $G[Y]$. But in $G^*$, every $T'$-vertex is adjacent to some vertices in $X'=X\setminus C'$ and
thus can be of degree $> d$. So Condition (1) may not hold. We will move some vertices out of $C'$ and $I'$ to
make the decomposition satisfying Condition (1).

Let $B$ be the set of $T'$-vertices that are of degree $>d$ in $G^*$.
Note that any vertex in $B$ is adjacent to some vertices in $X$.
We call vertices in $N_{I'}(B)=N(B)\cap I'$ \emph{bad} vertices.
Note that $B$ is not an empty set if and only if $N_{I'}(B)$ is not an empty set.
If $B=\emptyset$, then Condition (1) holds directly. For the case that $B\neq \emptyset$, i.e., $N_{I'}(B)\neq\emptyset$,
our idea is to update $I'$ by removing $N_{I'}(B)$ out of $I'$.
However, after moving some vertices out of $I'$, there may not be a full $(d+1)$-star packing from $C'$ to $I'$ anymore.
So after moving $N_{I'}(B)$ out of $I'$ we invoke the algorithm  ${\tt basic}(G[C'\cup I'],C',I')$  for Phase~2 on the subgraph $G[C'\cup I']$ to find new $C'$ and $I'$, and then check whether there are new bad vertices or not.
We do these iteratively until we find a $d$-bounded decomposition, where no bad vertex exists.
In the returned $d$-bounded decomposition, $I'$ and $C'$ may become empty.
However, we can guarantee $I'\neq \emptyset$ when the size of the graph satisfies some conditions.
We analyze this after describing the whole algorithm.

\medskip
\noindent
\textbf{The whole algorithm for decomposition.}
Our algorithm ${\tt decomposition}(G)$ presented in Figure~\ref{kernel} is to compute two subsets of vertices $C$ and $I$ of the input graph $G$
such that $(I, C, T=N(I)\setminus C, J=V\setminus(I\cup C\cup T))$ is a $d$-bounded decomposition of $G$.

Steps 3, 4 and 6 in ${\tt decomposition}(G)$ are the same steps in ${\tt basic}(G,X,Y)$. Here we add Step~5 into these steps, which is used to update
the $(d + 1)$-star packing $S$. In ${\tt decomposition}(G)$, Steps 1, 2 and 5 are corresponding to Phase~1, Steps 3, 4 and 6 are corresponding to Phase~2, and Steps 7 and 8 are corresponding to Phase~3. Note that Step~8 will also invoke ${\tt basic}(G,X,Y)$.

\begin{figure*}

\rule{\linewidth}{0.4mm}

\textbf{Input}: A graph $G=(V,E)$. \\
\textbf{Output}: Two subsets of vertices $C$ and $I$ such that $(I, C, T=N(I)\setminus C,J=V\setminus(I\cup C\cup T))$ is a $d$-bounded decomposition.
\begin{enumerate}
\item Find a maximal $(d + 1)$-star packing $S$ in $G$.
\item $X \leftarrow V(S)$ and $Y \leftarrow V\setminus X$.
\item Compute the bipartite graph $H$ and the auxiliary bipartite graph $H'$.
\item Compute a maximum matching $M'$ in $H'$ and the corresponding edge set $M$ and the $_{\leq} (d+1)$-star packing $S_{\leq d+1}$ in $H$.
\item Let $S_{d+1}$ be the set of $(d+1)$-stars in $S_{\leq d+1}$.\\
 \textbf{If} \{$|S_{d+1}|> |S|$\}, \\
 \textbf{then} $S \leftarrow S_{d+1}$ and \textbf{goto} Step 2.
\item Let $C'$ be $\emptyset$ if there is no untagged vertex, and be the set of $X$-vertices connected with at least one untagged vertex by an $M$-alternating path in $H$ otherwise. Let $X'\leftarrow X\setminus C'$. Let $Y'$ be the set of leaves of $_{\leq} (d+1)$-stars in $S_{\leq d+1}$ centered at vertices in $X'$ and let $I'\leftarrow Y\setminus Y'$.
\item Compute the set $N_{I'}(B)$ of bad vertices based on $C'$ and $I'$.
\item \textbf{If} \{$N_{I'}(B)\neq\emptyset$\}, \\
\textbf{then} $I' \leftarrow I'\setminus N_{I'}(B)$, $(C',I')\leftarrow {\tt basic}(G[C'\cup I'],C',I')$, and \textbf{goto} Step~7.
\item \textbf{Return} $(C=C',I=I')$.

\end{enumerate}

\rule{\linewidth}{0.4mm}
\caption{Algorithm ${\tt decomposition}(G)$}\label{kernel}
\end{figure*}

\llem{correct}{The two vertex sets $C$ and $I$ returned by ${\tt decomposition}(G)$
make $(I, C, T=N(I)\setminus C,J=V\setminus(I\cup C\cup T))$ a $d$-bounded decomposition.}
\pff{
To prove this  we only need to show the three conditions in the definition of $d$-bounded decomposition.
\refl{112} shows that the initial $C'$ and $I'$  satisfy Basic Condition. In Step~8, we will update $C'$ and $I'$ by taking a subset of each of them.
It is clear that there is a full $(d+1)$-star packing from $C'$ to $I'$ after updating them in Step 8, because we still use ${\tt basic}$ to compute new $C'$ and $I'$.
There is no edge between $I'$ and $X\setminus C'$ after each execution of Step~8, since \refl{112} guarantees that the vertices moved out of $X'$ in Step 8 are not adjacent to any vertices in the current $I'$. Then $C'$ and $I'$ in the whole algorithm always satisfy Basic Condition. Only when $N_{I'}(B)=\emptyset$, i.e., $B=\emptyset$, the algorithm will not execute Steps 7 and 8 anymore and stop. So when the algorithm stops, the decomposition based on $C=C'$ and $I=I'$ satisfy all the three conditions in the definition of $d$-bounded decomposition.
}

\begin{figure}[h]
\begin{center}
\includegraphics[width=0.9\textwidth]{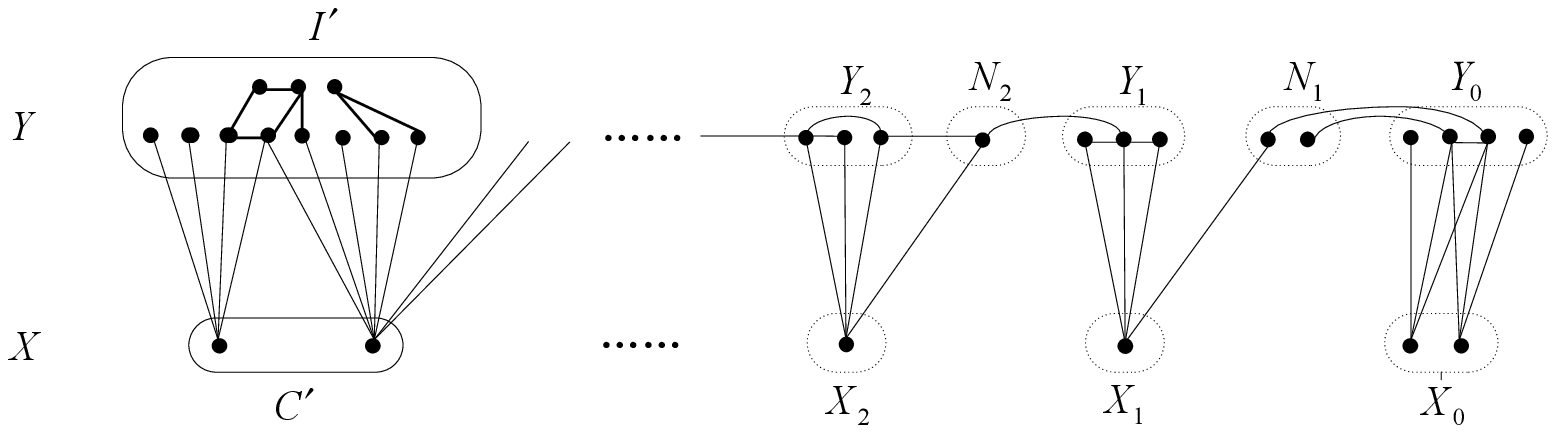}
\end{center}
\caption{An illustration for how ${\tt decomposition}(G)$ works, where we use $X_0$ (resp., $Y_0$)
to denote $X'$ (resp., $Y'$) computed in Step~6, $N_i$ to denote the set of vertices moved out of $I'$ in the  $i$th execution of $I' \leftarrow I'\setminus N_{I'}(B)$ in Step 8, and $X_i$ (resp., $Y_i$) to denote the
set of vertices moved out of $C'$ (resp., $I'$) in the
$i$th execution of $(C',I')\leftarrow {\tt basic}(G[C'\cup I'],C',I')$ in Step 8 for each $i\geq1$}\label{fig3}
\end{figure}

Figure~\ref{fig3} illustrates how the algorithm computes.
Next we consider the running time bound of the algorithm and show that it always stops.

Steps 1 and 2 take only linear time. We have analyzed in ${\tt basic}(G,X,Y)$ that Steps 3 and 6  take linear time and Step 4 uses $O(n^{1/2}m)$ time.
Each time when we update $S$ in Step 5, the size of $S$ increases by at least 1 and the size of $S$ is at most $\alpha(G)$ since each $(d+1)$-star contains
at least one vertex in a $d$-degree deletion set. Therefore, $S$ will be updated by at most $\alpha(G)$ times and the first six steps of
${\tt decomposition}(G)$ use $O(\alpha(G)n^{1/2}m)$ time.

Step 7 takes linear time. When $N_{I'}(B)\neq \emptyset$, Step 8 first moves some vertices out of $I'$ in linear time and
then updates $C'$ and $I'$ by calling ${\tt basic}(G,X,Y)$ in  $O(n^{1/2}m)$ time. We are interested in how many times Steps 7 and 8 will be executed.

For the purpose of presentation, we rewrite the second line of Step 8 as follows by using different notation:
\[\mbox{\textbf{then} $I'_0 \leftarrow I'\setminus N_{I'}(B)$, $(C^*,I^*)\leftarrow {\tt basic}(G[C'\cup I'_0],C',I'_0)$, and \textbf{goto} Step~7.}\]
Each time when execute Step 8, we have that
$$C^* \subseteq C', ~~I^*\subseteq I'_0\subseteq I' ~~~\mbox{and}~~~ N(I'_0) \setminus C' \subseteq I'\setminus I'_0.$$
First we consider the case that $C^* = C'$. Now we have that
$$N(I^*)\setminus C^* = N(I^*)\setminus C' \subseteq N(I'_0) \setminus C' \subseteq I'\setminus I'_0.$$
Each vertex in $I'$ is of degree at most $d$ in $G[V\setminus C']$ by \refl{112}.
So any vertex in  $N(I^*)\setminus C^*$ is of degree at most $d$ in $G[V\setminus C^*]$, which means that there is no bad vertex.
We conclude that: if $C^* = C'$, then $N_{I'}(B)$ will be empty in the next step and Step 8 will not be
executed any more.

By this property, we know that only when the size of $C'$ decreases the algorithm is possibly to execute the next iteration of Steps 7 and 8.
Initially, $|C'|\leq \alpha(G)$ since each $(d+1)$-star contains
at least one vertex in a $d$-degree deletion set. Therefore, Steps 7,8 and 9 of ${\tt decomposition}(G)$ run in $O(\alpha(G)n^{1/2}m)$ time.

In total, ${\tt decomposition}(G)$ uses $O(\alpha(G)n^{1/2}m)=O(n^{3/2}m)$ time.

\llem{time}{Algorithm ${\tt decomposition}(G)$ runs in $O(n^{3/2}m)$ time and returns $(C,I)$
such that $(I, C, T, J)$ is a $d$-bounded decomposition of $G$,
where $ T=N(I)\setminus C$ and $J=V(G)\setminus(I\cup C\cup T)$.
}

\refl{time} is not enough to prove \reft{normal_dc}, because $C$ and $I$ returned by ${\tt decomposition}(G)$ may be empty sets.
We still need to show that $I$ will not be empty if the size of the graph $G$ is large (compared to $\alpha(G)$).

\medskip
We prove the following lemma to show the size condition.
\llem{size}{Algorithm ${\tt decomposition}(G)$ returns $(C,I)$ such that
$$|V\setminus (C\cup I)| \leq (d^3+4d^2+5d+3)\alpha (G).$$
}
\pff{
%
After Step 5, $S$ will not be updated anymore. In our algorithm, we assume that $S$ is the one after Step~5 and will not change anymore. Note that
$C'$ and $I'$ are created and updated only after Step 5.

We let $s$ denote the number of $(d+1)$-stars in $S$. Then $s\leq \alpha(G)$ and $|X|=(d+2)s$.
Recall that $S_{\leq d+1}$ is a $_{\leq} (d+1)$-star packing from $X$ to $Y$ computed in Step~4.
Let $s_0$ be the number of $(d+1)$-stars in $S_{\leq d+1}$. Now we have $s_0\leq s$, otherwise $S$ would have been updated in Step~5.

In Step 6, initially $Y'$ is the set of leaves of  $_{\leq} (d+1)$-stars in $S_{\leq d+1}$ centered at vertices in $X'$. We let $Y_0=Y'$ in this step. Let $r_1$ be the number of $(d+1)$-stars in $S_{\leq d+1}$ centered at
some vertex in $X'$ and $r_2$ be the number of other stars in $S_{\leq d+1}$ centered at a vertex in $X'$. Then we have that $r_1+r_2\leq |X'|$ and $|Y_0|=|Y'|\leq (d+1)r_1 +dr_2$.
This is not the finial size of $Y'$, since some vertices more may be included to $Y'$ in Step~8.
Let $c_1$ denote the size of $C'$ in Step 6. Then we have that $c_1+r_1=s_0\leq s$ and $c_1+r_1+r_2\leq |X|$.

We consider the first execution of Step~8.
If $N_{I'}(B)\neq \emptyset$, then vertices in $N_{I'}(B)$ will be moved out of $I'$ and then will be included to $Y'$. Note that each vertex has degree at most $d$ in $G[Y]$, $B\subseteq Y'\subseteq Y$ and  $N_{I'}(B)\subseteq Y$. Then at most $|N_{I'}(B)|\leq d|B|\leq d|Y'|\leq d(d+1)r_1 +d^2r_2$ vertices will be moved out of $I'$.
So after executing $I' \leftarrow  I'\setminus N_{I'}(B)$ in Step 8 for the first time, the number of $Y$-vertices not in $I'$ is  at most
\[\begin{array}{*{20}{l}}
&|Y_0|+|N_{I'}(B)|\leq (d+1)r_1 +dr_2+(d(d+1)r_1 +d^2r_2)\\
=&(d+1)^2r_1 +d(d+1)r_2.
\end{array}\]
Now we have not analyzed the first execution of $(C',I')\leftarrow {\tt basic}(G[C'\cup I'],C',I')$ in Step~8 yet.

For each $i\geq 1$, assume that $x_i$ vertices are moved out of $C'$ in the $i$th execution of $(C',I')\leftarrow {\tt basic}(G[C'\cup I'],C',I')$ in Step 8. Then at most $(d+1)x_i$ vertices, the set of which is denoted by $Y_i$,
are moved out of $I'$ in this operation. In the $(i+1)$th execution of $I' \leftarrow I'\setminus N_{I'}(B)$, at most $d(d+1)x_i$ vertices will be moved out of $I'$ since $N_{I'}(B) \subseteq N(Y_i)\cap I' \subseteq N(Y_i)\cap Y$.
Note that if the algorithm executes Step~8 only for $i$ iterations, then we simply assume that $0$ vertices will be moved out of $I'$ in the $(i+1)$th iteration. In these two operations -- the $i$th execution of $(C',I')\leftarrow {\tt basic}(G[C'\cup I'],C',I')$ and the $(i+1)$th execution of $I' \leftarrow I'\setminus N_{I'}(B)$,
at most  $(d+1)^2x_i$ vertices are moved out of $I'$.

Finally, the number of $Y$-vertices not in $I=I'$ is at most
$$
y\leq(d+1)^2r_1 +d(d+1)r_2 +\sum_i{(d+1)^2x_i}.
$$
Note that $c_1=\sum_i x_i+|C|$, $c_1+r_1\leq s={\frac{|X|}{d+2}}$ and $r_1+r_2+c_1\leq |X|$.
We have that

\[\begin{array}{*{20}{l}}
y& \le &(d+1)^2(r_1+c_1-|C|)+d(d+1)(|X|-r_1-c_1)\\
{}& \le & (d+1)^2{\frac{|X|}{d+2}}+ d(d+1)|X|\\
& = & {\frac{(d+1)(d^2+3d+1)}{d+2}}|X|.
\end{array}\]
This inequality can be used to prove \reft{normal_dc}.

The number of $X$-vertices not in $C=C'$ is $|X|-|C|$. By $|X|=(d+2)s\leq (d+2)\alpha(G)$, we have

\[\begin{array}{*{20}{l}}
|V\setminus (C\cup I)| & = & |X|-|C|+y\\
{}& \le & {\frac{(d+1)(d^2+3d+1)}{d+2}}|X|+|X|\\
  & \le & (d^3+4d^2+5d+3)\alpha (G). \\
\end{array}\]
}

\refl{time} and the proof in \refl{size} imply \reft{normal_dc}.
The set $X$ after Step 5 in ${\tt decomposition}(G)$ is the special $d$-degree deletion set in \reft{normal_dc}.
So ${\tt decomposition}(G)$ finds the special $d$-degree deletion set and
$d$-bounded decomposition in \reft{normal_dc} in $O(n^{3/2}m)$ time.

\subsection{The algorithm for \reft{our-thm}}
Neither \reft{normal_dc} nor \refl{size} can get the size condition in \reft{our-thm} directly.
We use the following algorithm in Figure~\ref{bdd} for \reft{our-thm}.

\begin{figure*}

\rule{\linewidth}{0.4mm}

\textbf{Input}: A graph $G=(V,E)$. \\
\textbf{Output}: Two subsets of vertices $C$ and $I$ satisfying the conditions in \reft{our-thm}.
\begin{enumerate}
\item $C, I \leftarrow \emptyset$.
\item \textbf{Do} \{ $(C',I') \leftarrow {\tt decomposition}(G[V\setminus (C\cup I)])$, $C \leftarrow C\cup C'$ and $I \leftarrow I\cup I'$ \}\\
\textbf{while} $I'\neq \emptyset$.
\item \textbf{Return} $(C,I)$.
\end{enumerate}

\rule{\linewidth}{0.4mm}
\caption{Algorithm ${\tt BDD}(G)$}\label{bdd}
\end{figure*}

From the second iteration of Step~2 in ${\tt BDD}(G)$, each execution of $I \leftarrow I\cup I'$ will include at least  one new vertex to $I$.
So ${\tt decomposition}(G[V\setminus (C\cup I)])$ will be called for at most $n+1$ times.
Algorithm ${\tt BDD}(G)$ runs in $O(n^{5/2}m)$ time. Furthermore, if ${\tt decomposition}(G'=G[V\setminus (C\cup I)])$ returns
two empty sets, then by \refl{size} we have $|V(G')|=|V(G')\setminus (C\cup I)| \leq (d^3+4d^2+5d+3)\alpha (G')$. These together with
\refl{time} and \refl{size} imply \reft{our-thm}.

\section{Concluding Remarks}
In this paper, we provide a refined version of the generalized Nemhauser-Trotter-Theorem,
which applies to \textsc{Bounded-Degree Vertex Deletion} and for any  $d\geq 0$ can get a linear-vertex problem kernel for the problem parameterized by the solution size.
This is the first linear-vertex kernel for the case that $d\geq 3$.
Our algorithms and proofs are based on extremal
combinatorial arguments, while the original NT-Theorem uses linear programming relaxations~\cite{NT-theorem}.
It seems no way to generalize the linear programming relaxations used for the original NT-Theorem  to
\textsc{Bounded-Degree Vertex Deletion}~\cite{FG:gNT}.
A crucial technique in this paper is the
$d$-bounded decomposition.
To find such kinds of decompositions, we follow the ideas to find crown decompositions~\cite{A:crown2}
and the algorithmic strategy in~\cite{FG:gNT}. However, we use more ticks and can finally obtain the linear
size condition.

As pointed out by Fellows et al.~\cite{FG:gNT}, the results for \textsc{Bounded-Degree Vertex Deletion} in this paper can be modified for the problem of packing stars.
We believe that the new decomposition technique can be used to get local optimization properties and kernels for more deletion and packing problems.

Our algorithm obtains a kernel of $O(d^3k)$ vertices for \textsc{Bounded-Degree Vertex Deletion} when $d$ is also part of the input. Another interesting problem for further study
is to investigate the lower bound of the kernel size for the dependency on $d$.

\section*{Acknowledgement}
The author was supported in part by National Natural Science Foundation of China under the Grant
61370071.

\end{document}